# Information and Multi-Sensor Coordination

Greg Hager and Hugh Durrant-Whyte*
GRASP Laboratory
Department of Computer and information Science
University of Pennsylvania
Philadelphia, PA 19104

## Abstract

The control and integration of distributed, multi-sensor perceptual systems is a complex and challenging problem. The observations or opinions of different sensors are often disparate, incomparable and are usually only partial views. Sensor information is inherently uncertain, and in addition the individual sensors may themselves be in error with respect to the system as a whole. The successful operation of a multi-sensor system must account for this uncertainty and provide for the aggregation of disparate information in an intelligent and robust manner.

We consider the sensors of a multi-sensor system to be members or agents of a team, able to offer opinions and bargain in group decisions. We will analyze the coordination and control of this structure using a theory of team decision making. We present some new analytic results on multi-sensor aggregation and detail a simulation which we use to investigate our ideas. This simulation provides a basis for the analysis of complex agent structures cooperating in the presence of uncertainty. The results of this study are discussed with reference to multi-sensor robot systems, distributed AI and decision making under uncertainty.

## 1 Introduction

The general problem of seeking, sensing, and using perceptual information is a complex and, as yet, unsolved problem. Complications arise due to inherent uncertainty of information from perceptual sources, incompleteness of information from partial views, and questions of deployment, coordination and fusion of multiple data sources. Yet another dimension of complexity results from organizational and computational considerations. We feel that these three topics – information, control, and organization – are fundamental for understanding and constructing complex, intelligent robotics systems. In this paper, we are concerned with developing useful analytic methods for describing, analyzing and comparing the behavior of such constructions based on these criteria.

*This material is based on work supported under a National Science Foundation Graduate Fellowship and by the National Science Foundation under Grants DMC-8411879 and DMC-12838. Any opinions, findings, and conclusions or recommendations expressed in this publication are those of the authors and do not necessarily reflect the views of the National Science Foundation.

We assume from the outset that such robotic systems are basically task-oriented, goal-directed agents. The behavior of a system is determined entirely by the goal it is working toward, and the information it has about its environment. At any point in time, such an agent should use the available information to select some feasible action. The most preferable action should be that which is expected to lead the system closest to the current goal. In short, we will consider the question of driving robotics systems as a large and complex problem in estimation and control. To adopt the nomenclature of decision theory[2], at any point in time an agent has a local *information structure* reflecting the state of the world, a set of *feasible actions* to choose from, and a *utility* which supplies a preference ordering of actions with respect to states of the world. We generally assume that a *rational* decision maker is one which, at any point in time, takes that action which maximizes his utility. Our commitment, as a result of casting the problem in a decision theoretic perspective, is to provide principled means for specifying information structures, actions, and (perhaps most crucially) determination of utility.

This monolithic formulation is certainly too naive and general to successfully attack the problem. The state of the system is a complex entity which must be decomposed and analyzed to be understood. The resulting procedures for control will undoubtedly be computationally complex. Computer resources, like human problem solvers, have resource limitations which bound the complexity of problems that can be solved by a single agent – otherwise known as *bounded rationality* [25]. Such computational considerations suggest distributing the workload to increase the problem solving potential of the system. From a practical standpoint, the system itself is composed of physically distinct devices, each with its own special characteristics. Software and hardware modules should be designed so that information and control local to subtasks is kept locally, and only information germane to other subtasks is made available. Ultimately, sensors and subtasks could independent modules which can be added or removed from a system without catastrophic results. In this case we desire each subtask to have the ability to cooperate and coordinate its actions with a group while maintaining its own local processing intelligence, local control variables, and possibly some local autonomy.

Our solution is to view the systems as decomposed into several distinct decision-makers. These modules are to be organized and communicate in such a manner as to achieve the common goal of the system. Organizations of this type of are often referred to as teams[8,16]. We propose to consider a team theoretic formulation formulation of multi-sensor systems in the following sense: The agents are considered as members of the team, each observing the environment and making local decisions based on the information available to them. A manager (executive or coordinator) makes use of utility considerations to converge the opinions of the sensor systems. Section 2 will be devoted to a review of team decision theory and present some new analytic results[7].

One criticism of decision theory is that optimal solutions are often difficult or impossible to find. In order to aid in analysis of these problems, we have built a simulation environment. We use the simulation to examine various non-optimal and heuristic solutions to otherwise intractable problems, and experiment with different loss functions to determine the character of the resultant decision method. The simulation is a generalization of classic pursuit and evasion games [14] to teams of pursuers and evaders. Each team member has local sensors and state variables. They are coordinated through a team executive. Section 3 will be devoted to a detailed look at the simulation and our results to date.

We feel that the team formulation of sensory systems has implications for the broader study of Artificial Intelligence. AI is relevant to this work in at least two respects:

- Firstly, it is certainly possible to consider the agents of the system as performing some reasoning process. Considering AI systems as decision-makers seems a plausible approach to the construction of intelligent distributed systems. Thus, this work has commonalities with Distributed AI in that both are interested in questions of structuring information and communication between intelligent systems.

- Secondly, we often want to *interpret* the information available to the system, and to communicate information as interpretations rather than simple signals. This is primarily a problem in representation of information. Again, AI has focussed on the interpretation of information, and the representation of that interpretation.

More generally, we would like to discover when systems like this can be profitably posed as decision problems. Section 4 will be devoted to an in depth discussion of the general merits and shortcomings of the organizational view, and attempt to define when it is most appropriate or useful.

## 2 A Team-Theoretic Formulation of Multi-Sensor Systems

Team theory originated from problems in game theory [26] and multi-person control. The basis for the analysis of cooperation amongst structures with different opinions or interests was formulated by Nash [20] in the well known bargaining problem. Nash's solution for the two person cooperative game was developed into the concepts of information, group rationality and multi-person decisions by Savage [24]. Team theory has since been extensively used by economists to analyze structure [16], information [18] and communication. Section 2.1 introduces the team structure and defines the function of the team members and manager. Different team organizations are discussed and the concepts of information structure, team decision, team utility and cooperation are defined in Section 2.2. Section 2.3 applies these techniques to the multi-sensor team and a method for aggregating opinions is derived. Due to lack of space, we will assume some familiarity with probability and decision theory.[1]

### 2.1 Team Preliminaries

A sensor or member of a team of sensors is characterized by its information structure and its decision function. Consider a team comprising of $n$ members or sensors each making observations of the state of the environment. The information structure of the $i^{th}$ team member is a function $\eta_i$ which describes the character of the sensor observations $z_i \in \mathcal{X}_i$ in terms of the state of the environment $\theta \in \Theta$, and the other sensor actions $a_j \in \mathcal{A}_j$, $j = 1, \cdots, n$. So that:

$$z_i = \eta_i(\theta, a_1, \cdots, a_n) \qquad (1)$$

Collectively the n-tuple $\eta = (\eta_i, \cdots, \eta_n)$ is called the information structure of the team. The action $a_i$ of the $i^{th}$ team member is related to its information $z_i$ by a decision function $\delta_i \in \mathcal{D}_i$ as $a_i = \delta_i(z_i)$. We may also allow randomized rules, in which case $\delta$ associates information with a *distribution* over the set of feasible actions. Collectively the n-tuple $\delta = (\delta_1, \cdots, \delta_n)$ is called the team decision function. For an estimation problem, the action space $\mathcal{A}_i$ is the same as the space of possible states of nature $\Theta$: Our action is to choose an estimate $a_i = \theta_i \in \Theta$.

There are a number of different forms that the information structure can take which in turn characterizes the type of problem to be solved. If for all team members $\eta_i$ is defined *only* on $\Theta$ ($\eta_i: \Theta \longrightarrow \mathcal{X}$) the resulting structure is called a *static* team [16]. When $\eta_i$ also depends on the other team members' actions, then the structure is called a *dynamic team* [13]. Clearly as each team member can not make decisions and be aware of the result simultaneously, the general form of information structure for a dynamic team must induce a causal relation on the team member actions $a_j$. We can apply a precedence structure on the time instant a member makes a decision, so that if member $i$ makes a decision prior to member $j$ then the information structure $\eta_i$ will not be a function of $a_j$. Indexing the team members by their decision making precedence order we can rewrite the information structure as:

$$z_i = \eta_i(\theta, a_1 \cdots, a_{i-1})$$

---

[1] An extended version of this paper appears as Grasp Lab tech. report 71.

A sensor or member of a team will be considered rational if it can place a preference ordering on its actions that admits a utility function $u_i \in U$. One possible set of rationality axioms can be found in [2, p. 43] and the proof that these axioms admit a utility function can be found in [5]. A decision rule $\delta(\cdot)$ can be evaluated in terms of its payoff:

$$u_i(\delta, \theta) = \int_{\mathcal{X}_i} u_i(\delta(z_i), \theta) f(z_i|\theta) dz_i = E[u_i(\delta(z_i), \theta)]$$

We assume that a rational team member is attempting to maximize its payoff.

The team utility is a function which assigns a value to each team action $L(\theta, a_1, a_2, \cdots, a_n)$. The role of $L$ is very important in characterizing the team. The interpretation of team action due to Ho, Chu, Marschak and Radnor [13,16], is that the goal of every team member is to maximize $L$ regardless of personal loss (in fact, personal loss is not even defined). We will call this an "altruistic" team. An alternative formulation is to allow individual team members to have a personal utility as well as an interest in the team. For example a team member may agree to cooperate and be subject to the utility $L$, or to disagree with the other team members and be subject to a personal utility. In this case a rational team member will agree to cooperate *only* if it will gain by doing so: when the team utility exceeds its personal utility. We shall call this an "antagonistic" team.

The idea of individual rationality can be extended to include so-called group rationality. Nash first introduced a set of group rationality axioms. There has been considerable disagreement about these axioms [28], and a number of other definitions have been suggested *e.g.* [10]. The underlying basis for providing group rationality is the ability of a team to put a preference ordering on group decisions. Unlike individual utility considerations, this involves a number of assumptions about the nature of the group or team. For example, each team member must assume some subjective knowledge of other players rationality, interpersonal comparisons of utility require preferences to be congruent and assumptions must be made about indifference, dominance and dictatorship.

## 2.2 Team Organizations

The problems associated with the extension of individual to group rationality are all concerned with the comparison of individual utilities. The existence of a group preference ordering is equivalent to requiring that the combination of individual team member utilities that form the team utility, is convex. If this is satisfied then we say that the group decision is also person-by-person optimal. The key principle in group decision making is the idea of Pareto optimal decision rules:

*Definition*: The group decision $\delta^*$ is Pareto-optimal if every other rule $\delta \in D$ decreases *at least* one team members utility.

If the risk set of the team $L(\theta; \delta_1, \cdots, \delta_n) \in \mathcal{R}^n$ is convex, then it can be shown [13] that such a team decision is also person-by-person optimal so that for all team members $i = 1, \cdots, n$ the team action $\mathbf{a} = [a_1, \cdots, a_n]^T$ also satisfies

$$\max_{a_i \in A_i} E\left[L(\delta_1^*(z_1), \cdots, a_i = \delta_i(z_i), \cdots, \delta_n^*(z_n))\right] \quad (2)$$

If the class of group decision rules $D$ includes all jointly randomized rules then $L$ will *always* be convex. If we really believed in an altruistic team, we must use this class and be subject to these results. Considerable work has been done on finding solutions to equation 2.3 under these conditions [16,13,12,11], particularly as regards the effect of information structure on distributed control problems.

We are primarily interested in teams of observers – sensors making observations of the state of the environment. In this case the team members can be considered as Bayesian estimators, and the team decision is to come to a consensus view of the observed state of nature. The static team of estimators is often called a Multi-Bayesian system [28]. These systems have many of the same characteristics as more general team decision problems. Weerahandi [27] has shown that the set of non-randomized decision rules is not complete in these systems. If two team members using decision rules $\delta = [\delta_1, \delta_2]$ have utilities $\mathbf{u}(\theta) = u_1(\delta_1, \theta)$ and $u_2(\delta_2, \theta)$, then the team utility function $L(\theta) = L(\mathbf{u}(\theta))$ will only admit a consensus if it satisfies the inequality:

$$E[L(\mathbf{u}(\theta))] \geq L(E[\mathbf{u}(\theta)]) \quad (3)$$

This is the Jensen inequality, and it is well known that this will be satisfied if and only if the function $L(\mathbf{u}(\theta))$ and the risk set are convex. Generally, this will only be true when the set $D$ of decision rules includes jointly randomized decision rules.

Consider the team utility $L$ as a function of the team member utilities so that $L = L(u_1, \cdots, u_n) = L(\mathbf{u})$. The group rationality principles described above restrict the functions $L$ that are of interest to those that have the following properties[1]:

1. Unanimity: $\frac{\partial L}{\partial u_i} > 0, \forall i$

2. No dictator: If $\forall i: u_i \neq 0$, there is no $u_j$ such that $L = u_j$.

3. Indifference: If $\forall i, \exists \delta_1, \delta_2$ such that $u_i(\delta_1, \cdot) = u_i(\delta_2, \cdot)$, then $L(\delta_1) = L(\delta_2)$

If the team utility function $L$ satisfies these properties, we will say that the team is rational. The function $L$ is often called an "opinion pool". Two common examples of opinion pools are the generalized Nash product:

$$L(\theta; \delta_1, \cdots, \delta_n) = c \prod_{i=1}^{n} u_i^{\alpha_i}(\delta_i, \theta) \quad \alpha_i \geq 0$$

and the logarithmic or linear opinion pool:

$$L(\theta; \delta_1, \cdots, \delta_n) = \sum_{i=1}^{n} \lambda_i u_i(\delta_i, \theta), \quad \lambda_i \geq 0, \quad \sum_{i=1}^{n} \lambda_i = 1$$

The value of the generalized Nash product can be seen by noting that if $u_i(\delta_i(z_i),\theta) = f(z_i \mid \theta)$ and $\alpha_i = 1$ then $L$ is the posterior density of $\theta$ with respect to the observations $z_i$. A criticism leveled at the generalized Nash product is that it assumes independence of opinions, however this may be accounted for through the weights $\alpha_i$. A criticism of the linear opinion pool is that there is no reinforcement of opinion.

Suppose we now restrict group decision rules $\delta \in D$ to non-randomized decisions. This allows team members to disagree in the following sense: If the team risk set $\mathbf{u} = [u_1(\delta_1,\theta), \cdots u_n(\delta_n,\theta)]$ is convex for non-randomized $\delta$, then equation 3 holds and a consensus may be reached. If however $\mathbf{u}$ is concave in at least one $u_i$, and if randomized rules are disallowed, it is better (in terms of utility) for the associated team members to disagree: as if they were acting as an an antagonistic team. It should be clear from this example that the difference between antagonistic and altruistic teams is the ability to obtain a convex "opinion" space.

If all the $u_i$ are convex functions, then $L$ will always be convex on the class of non-randomized decisions. However in location estimation or Multi-Bayesian systems, the $u_i$ will often be concave so that $L(\mathbf{u})$ will be guaranteed convex only in the class of randomized rules. Thus $L(\mathbf{u})$ will always be convex for an altruistic team. For an antagonistic team $L$ will only be convex when agreement can be reached (in the class of non-randomized decisions), otherwise if opinions diverge sufficiently then $L$ will be concave. Concavity will generally take the form of separating team members into convex groups of opinions coalitions which may overlap.

Our interest in these results centers on finding when agreement can be reached and in calculating the value of the consensus. We summarize these concepts in the following:

*Result 1*: Consider a team with member utilities $u_i(\delta_i,\theta)$ and team utility satisfying the group rationality conditions. Then:

1.1. *Consensus*: Cooperation will only occur when the set of risk points $L(\delta_1, \cdots, \delta_n) \in \mathcal{R}^n$ is convex.

1.2. *Altruistic*: If $\delta \in D$ is the class of all randomized decision rules then $L$ will always be convex.

1.3. *Antagonistic*: If $\forall i, L \geq u_i$ then L will be convex in the class of non-randomized decision rules.

1.4. *Disagreement*: When $L$ is concave there is no best decision and agreement cannot be reached.

The point at which $L$ becomes concave for each member is called the disagreement point, the value of a member's utility at this point is called the security level.

### 2.3 Multi-Sensor Teams

The fusion of sensor observations requires that we have a method for comparing information from disparate sources. We consider each sensor to be a member of an antagonistic team in the following sense: Each sensor comes up with uncertain partial views of the state of the environment, the goal of the executive is to integrate the various sensor opinions, by offering incentives and interpretations for combining disparate viewpoints. The antagonistic team structure allows members to disagree if for some reason they have made a mistake or cannot reconcile their views with those of the other team members. An altruistic team could not take this action.

We suggest that the comparison of diverse observations can be interpreted in terms of a comparison of the utility of a consensus decision. Suppose we have two observations $z_1$ and $z_2$ which are not directly comparable. Each observation contributes to some higher level description of the environment, and each is dependent on the other. We can interpret any decision $\delta$ about the environment in terms of its utility to the observations: $u_1(\delta(z_1),\theta)$ and $u_2(\delta(z_2),\theta)$. Although $z_1$ and $z_2$ cannot be compared directly, their contributions to particular decisions can be evaluated in a common utility framework. The team theoretic comparison of utilities admits a measure of disagreement and allows for the evaluation of sensor information in a consistent manner.

Define $\Theta$ to be the set of states of nature and consider a robot system with sensors $S_j$, $j = 1, \cdots, m$, taking sequences of observations $\mathbf{z}_i = \{z_i^1, \cdots, z_i^k\}$ of features in the environment. We will restrict interest to the static team structures so that $\mathbf{z}_i = \eta_i(\theta)$. Locally, sensors can make decisions based on local observations as $\theta = \delta_i(z_i)$ from comparable sequences $\mathbf{z}_i = \{z_i^1, \cdots, z_i^k\}$, with respect to a common utility $u_i(\delta_i(z_i),\theta)$. Jointly the sensor team has a utility $L = L(\theta;\delta_1, \cdots, \delta_n)$, which can be considered as a function of the individual utilities $L = L(u_1, \cdots, u_2)$ satisfying the group rationality conditions.

If the observations from different sensors are incomparable, they must be interpreted in some common framework. This will be the case when the sensors are located in different locations for example. Let $D_i$ interpret $S_i$'s observations in some common description framework. Then the team loss can be written as:

$$L(u_1(\delta_1[D_1(z_1)],\theta), \cdots, u_n(\delta[D_n(z_n)],\theta)])$$

By selecting $L$ and analyzing its convexity, we will establish the character of the sensor team.

The rationality axioms derived from utility theory require that we be able to put a preference ordering on decisions $\delta_j(\cdot)$. It seems reasonable that the preference ordering admitted by an observation $z_i$ will be the same ordering as that obtained by a maximum likelihood estimator (unbiased Bayes rationality). In this case, the utility function of an observation will be coincident with its likelihood function. Thus the Gaussian distribution $N(z_i,\Lambda_i)$ associated with the observation $z_i$ can also be considered as the preference ordering or posterior utility function of $z_i$ on any resulting estimate $\theta$. In this framework, two observations $z_i$ and $z_j$ will have a basis for agreement only if their combined utility exceeds their individual utility, that is a consensus can only be reached if the set of observation utilities form a convex set.

To fix this, define $u_i(z,\theta) = f_i(z_i|\theta) \sim N(z_i, \Lambda_i)$ as the loss to the observation $z_i$ of the estimate $\theta$, and let

$$L(\theta) = L(u(\theta)) = [u_1(z_1,\theta), \cdots, u_n(z_n,\theta)]^T$$

denote the team utility. Then, in terms of expected utility, a consensus can only be reached if $L$ satisfies Equation 3, *i.e* the function $L$ is convex.

The function $L$ will be convex if and only if its matrix of second order derivatives is non-negative definite. If $L$ satisfies the group rationality principles, this requires that $\frac{\partial^2 u_i}{\partial \theta^2} \geq 0$ for $i = 1, \cdots, n$. Differentiating $u_i$ gives

$$\frac{\partial^2 u_i}{\partial \theta^2} = \left[1 - (\theta - z_i)^T \Lambda_i^{-1} (\theta - z_i)\right] \cdot N(z_i, \Lambda_i)$$

For these to be positive, and hence the set $u \in \mathcal{R}^n$ to be convex, we are must find a $\theta$ which satisfies:

$$(\theta - z_i)^T \Lambda_i^{-1} (\theta - z_i) \leq 1 \quad (4)$$

For all $i = 1, \cdots, n$ observations.

Consider any two observations $z_i$ and $z_j$. They can form a consensus if we can find a $\theta$ that satisfies equation 4 for both $z_i$ and $z_j$. To compare observations, we interpret them in a common framework as $D_i(z_i)$ and $D_j(z_j)$. If $J_i$ and $J_j$ are the jacobians of $D_i$ and $D_j$ respectively [6], define ${}^D\Sigma_i = J_i^{-1} \Lambda_i^{-1} J_i^{-T}$. This is the information matrix of the observation $z_i$ transformed to the common frame of reference by the transformation $D$.

Since the left hand side of equation 4 is always positive, we must find a $\theta$ which satisfies

$$\frac{1}{2}(\theta - D_i(z_i))^T {}^D\Sigma_i (\theta - D_i(z_i)) +$$
$$\frac{1}{2}(\theta - D_j(z)_j)^T {}^D\Sigma_j (\theta - D_j(z_j)) \leq 1 \quad (5)$$

The value of $\theta$ which makes the left hand side of this equation a minimum (and which is also the consensus when it exists) is given by the usual combination of normal observations[2]:

$$\theta = \left({}^D\Sigma_i + {}^D\Sigma_j\right)^{-1} \left({}^D\Sigma_i D_i(z_i) + {}^D\Sigma_j D_j(z_j)\right)$$

Substituting this into equation 5 gives:

$$\frac{1}{2}(D_i(z_i) - D_j(z_j))^T {}^D\Sigma_i \left({}^D\Sigma_i + {}^D\Sigma_j\right)^{-1}$$
$${}^D\Sigma_j (D_i(z_i) - D_j(z_j)) \leq 1 \quad (6)$$

We will say that $z_i$ and $z_j$ admit a Bayesian (non-randomized) consensus if and only if they satisfy equation 6. The left side of Equation 6, which we will denote as $d_{ij}^2$, is called the generalized Mahalanobis distance (a restricted form of this is derived in [27]) and is a measure of disagreement between two observations. Figure 2.3 shows plots of $u_i$ against $u_j$ for various values of $d_{ij}^2$ and which clearly demonstrate that the convexity of the set $[u_i, u_j]$ corresponds to requiring that $d_{ij}^2 \leq 1$. This measure can be further extended to consider more than two observations at a time. For example, if each observation $z_i$, $i = 1, \cdots, n$ has

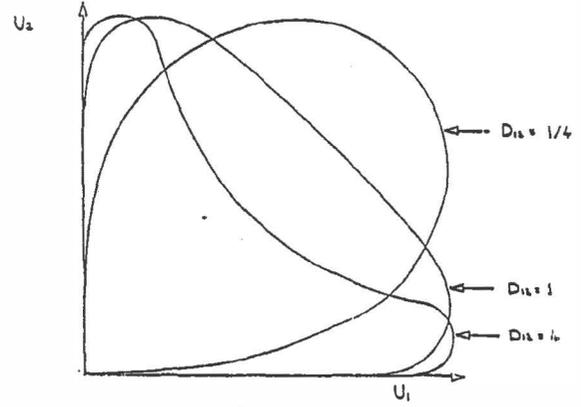

Figure 1: Plots of Mahalanobis distances

the same variance-covariance matrix $\Lambda$, then a consensus be obtained only if:

$$\frac{1}{2n^2} \sum_{i=1}^{n} \sum_{j=1}^{n} d_{ij} \leq 1 \quad (7)$$

It is clear that a set of observations that satisfy Equation 6 pair-wise will also satisfy Equation 7

In most real situations, it is unlikely that we will know the variance-covariance matrices exactly. In this case, any estimates of the $\Lambda_i$ act as if they were thresholds in the sense that the larger the $\Lambda_i$ that is used, the more disagreement will be tolerated.

## 3 Simulation Studies

To this point, we have discussed the theoretical aspects of estimation in the team framework. Our goal is to eventually pose problems of multi-sensor control in coordination and solve them in a similar manner. However, finding and analyzing solutions to decision, control, or game problems, especially in the face of anything less than perfect information, can be extremely difficult. From a technical perspective, solutions under even relatively simple losses are complex optimization problems. Other heuristic or *ad hoc* approaches must often be considered. Methodologically, there is a question as to what proper loss functions are for different problems. Ideally, the loss function should reflect the actual state of affairs under consideration since it reflects the preferences of the decision maker. Whereas in the economics literature, losses are usually derived from utility considerations based on monetary rewards, we have a much wider set of competing criteria to consider. This complicates matters to the point that we need to gain intuition about the issues involved before hypothesizing a solution.

In order to deal with these issues, we have constructed a simulation on a Symbolics Lisp Machine. The simulation takes the form of a game of pursuit and evasion similar in character to the classic differential game known as the *homocidal chauffeur* [14]. That classical form of this game consists of a pursuer and evader moving at constant velocity in a plane. Both players have perfect information about the other's state, and attempt use this information to intercept or evade their opponent respectively. The payoff structure of the game is the time until capture. The major changes we have made are that we have equipped the players with imperfect sensing devices (i.e. the players use imperfect state information), and we allow multiple pursuers and evaders grouped into teams coordinated by a team executive. It is important to note that the motivation for using the pursuit-evasion framework is primarily to provide each team with a well-defined method for comparing structures and control policies. The game is not of intrinsic value by itself, but forms a structured, flexible, closed system in which sensor models, organizational structures and decision methods may be implemented and easily evaluated.

The simulation is constructed so that we can vary the structure of team members, as well as overall team structure, and quickly evaluate the effects of the change based on the character of the simulated game that ensues. We have in mind to allow variation in such factors as dynamics, sensors, information integration policies, incentive structures, and uncertainty of information, and observe what types of policies lead to the adequate performance in these circumstances. We expect to transport what we learn from the simulation to real-world problems of multi-sensor robot systems currently being developed in the Grasp laboratory[21]. We imagine a situation where this simulation provides an environment in which distributed expert coordination and control problems can be investigated before implementation and conversely that applications of the sensor systems under development will suggest what directions, sensor models and dynamics would most fruitful to explore in the simulation. The remainder of this section details the current structure of the simulation environment, and outlines our initial experiences with it.

### 3.1 The Game Environment

The simulation takes place on a planar field possibly littered with obstacles. The basic cycle execution involves team members taking sensor readings, executives integrating information and offering incentives, and finally team members making decisions. The state variables are updated and the game moves to a new step. A game terminates when and if the pursuit robots, which are equipped with a simple ballistics system, capture all the evaders. This is a medium level of granularity with emphasis on the general behavior of teams, not the precise performance issues of team members. Some time-constraint issues can be investigated by including time parameters in the payoff functions, but computational complexity issues and investigations of asynchronous behavior are outside the scope of our considerations. For instance, if some decision policy is computationally more complex than another, differences in performance will not reflect that complexity.

### 3.2 The Structure of Team Members

The character of individual team members is determined by three modules:

1. The kinematics and dynamics of motion on the plane,

2. what sensors are available and the noise characteristics of those sensors, and their kinematics and dynamics, and

3. the ballistics which determine the termination of the game.

The team members are constant velocity, variable direction units operating in a plane with state variables $x$, $y$, and $\theta$. Since the robots move with constant velocity, the only directly controlled variable is $\theta$. The only dynamical consideration involved is how we allow the robot to change its current heading to some new desired heading $\theta_d$ – the single control. Currently, we assume that when reorienting each agent can move with some fixed (possibly infinite) velocity, $\omega$. The has the effect of defining a minimal turning radius. Pursuers generally have some finite $\omega$, while evaders have infinite $\omega$ – *i.e.* they turn instantaneously.

The sensor model we are currently using is a range and direction sensor. The sensor has a limited cone of data gathering, and a limited range. It has a single control variable $\alpha$ which the robot can select to point the sensor. We assume that sensors typically return noisy data, so we have different noise models which we "wrap around" the sensor to make it more closely approximate real data gathering devices. The induces decision problems in dealing with both the noise and range limitations of devices. The fact the the sensors are distributed introduces issues in integrating noisy observations from different frames of reference[6]. Finally, since sensors are transported by the robot, there are issues involved in resolving the conflicts between action for pursuit or evasion, and actions which will allow more efficient gathering of information.

Termination of the game occurs when all evaders are eliminated. We define a *capture region* which delineates how close a pursuer must come to eliminate an evader. However, when information is noisy, the area in which the evader can be located will have an associated uncertainty. We sometimes equip each pursuer with some mechanism to "shoot" evaders, allowing the possibility of uncertainty in observation to make it "miss". Part of the payoff structure of the game can include costs for using projectiles and missing; thereby adding incentive to localize the evader to the best degree possible.

### 3.3 Information Structures, Organization, and Control

The intesting issues are how the robot systems are controlled, and how team members interact. Each team member must make decisions, based on available information, about the best setting of its control variables. Thus, each team member has a local information structure and a utility function or decision rule as outlined in Section 2. The type of information available – its completeness and fidelity – along with the utility function determine how an agent behaves for a fixed set of controls. We have specifically modularized information structure and decision processes so that variations are easily compared.

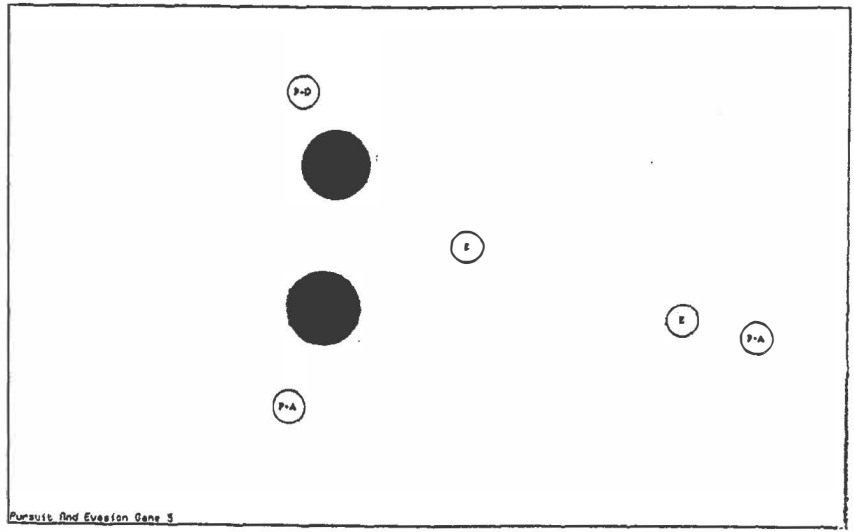

Figure 2: The Pursuit-Evasion Game

The information structure can consist of only local information, or can contain information communicated from other team members, as well as computed information based on observation history. The team executive provides the basic organizational structure of the team. It is equipped with the team information structure, and computes the team utility which is offered as the incentives for a team member to cooperate with the team. The issues here involve integrating the team information, and the exact nature of the team utility. In our experiments, we either use the executive as an information integrator and team blackboard, or restrict robots to their own local information.

Our experiments to date have been in controlling the direction of travel of robots. Three decision methods have been used:

1. Purely local information and decision

2. Completely global information and a linear pool team utility (see page 3)

3. A mix of global and local information with a non-convex utility structure.

The first of these is the obvious strategy of chasing whatever is within the viewing radius and avoiding any obstacles. The second amounts to the executive choosing an evader to follow, and the team members agreeing at all costs. This has the undesirable property of the pursuers being destroyed while attempting purely global objectives. The first method disregards centralized control, possibly missing an opportunity for team members to cooperate for common benefit. The latter disregards individual concerns for global objectives, possibly disregarding important local objectives.

The final method uses the executive to integrate information about evaders, and to offer a team incentive to chase a particular evader. But, it also lets team members compute an incentive to avoid obstacles which fall on their path. Figure 2 shows a team configuration where some team members (those labeled "P-D") are disagreeing with the team in order to avoid an obstacle, while the rest of the team (labeled "P-A") are following the executives order to chase an evader. This is our first experience with a mix of local and global control.

Our next objective in the simulation is to consider noisy observations and develop sensor control algorithms. Our idea for this project is the following: recall that sensors return distance and direction. We henceforth assume both quantities are distributed according to some probability distribution. The information structure of the team will consist the current best estimate and the information matrix of measurements integrated as in Section 2. The utility for an angle $\alpha_i$ of a sensor $i$ will be the expected change in the information for the closest evader within the cone of vision. This means that individual members will choose that evader for which they can contribute the "maximum" information. We have not developed any team policies for this scenario, yet.

## 4 Evaluation and Speculation

We have considered a very basic, static, non-recursive team structure for the sensors and cues of a robot system. The results obtained for the aggregation of agent opinions are intuitively appealing and computationally very simple. Similarly, the initial simulation experiments with distributed control seem promising. However, it is clearly the case that the methods presented thus far could easily be developed with-out recourse to the concepts of team and information structure. We have chosen to introduce these ideas for two main reasons: firstly as a device through which the interactions between sensor agents may be easily explained, and secondly because we feel that team theoretic methodology has great potential for understanding and implementing more complex organizational structures in a systematic manner. Our main point is that team theory is neither a completely abstract non-computational formalization of the problem, nor a computational technique or algorithm with no theoretical potential, but is in fact our analog of a *computational theory*[15]. We assert that the inherent elements of cooperation and uncertainty make team theory the appropriate tool for this class of problems[11]. This section discusses the advantages of team theory suggests issues which need to be explored more fully.

## 4.1 Information and Structure

Many of the advantages of team theoretic descriptions lies in the ability to analyze the effects of different agent (team member) information structures on the overall system capabilities. Recall that in the general case, the $i^{th}$ team members information structure may well depend on the other team members actions and information, either as a preference ordering (non-recursive) or in the form of a dialogue (recursive) structure. For example, consider a stereo camera and a tactile sensor, acting together in a team. It is often the case that the stereo matching algorithm is incapable of finding disparities and three dimensional locations that are horizontal in the view plane, whereas a tactile sensor mounted on a gripper jaw is *best* at finding just such horizontal features. In addition it is reasonable to assume that, while a vision system is good at finding global locations, a touch sensor is better for refining observations and resolving local ambiguities. What is required is a sharing of information between these two sensors: their respective information structures should be made dependent on each other's actions. We can imagine specifying the problem so that the solution is anything from a simple optimal control to an extended dialogue taking place between the two sensors, resolving each others observations and actions, arriving at a consensus decision about the environment. This example clearly shows the advantages of a team theoretic analysis. We can postulate alternative information structures for the sensors and the dynamics of the exchange of opinions can be analyzed: Is a consensus obtained? When is a decision made? Should communication bandwidth be increased or decreased? etc.

Another aspect of this scenario is that the two sensors have partitioned the environment into a kind of "who knows what" information structure. In general, not all the information about a robotics system is relevant to the construction of specific portions of the system. Analogously, all the information available via sensors is not relevant to the performance of all parts of the system. In the example above, the spatial characteristics of the camera image are of interest only to the camera agent, and the response characteristics of the tactile sensor are of relevant only to the tactile controller. The ambient illumination as measured by the camera has no relevance to decision made by the tactile sensors even though they may cooperate in disambiguating edges. Team theory allows information and control to reside *where it is appropriate*, and thereby reduces problem complexity and increases performance potential. We believe that this is a crucial principle for the construction of intelligent robotics systems.

To this point, we have not discussed uncertainty of information. However, information from perceptual sources is sure to have some associated uncertainty. Uncertainty adds an entire dimension to any discussion of information – we must consider some grade of belief in information[4] and how that should influence the choice of action. In the case of perfect sensing, information is either adequate or inadequate; and new information can be derived by using the constraints of the problem at hand. Hence, new facts will either lead to more information, or be redundant. On the other hand, if information is uncertain, adding more *unrelated* observations may not really increase the available information. Multiple correlated observations may, in fact, be a better strategy since that is likely to reduce uncertainty. Encoding considerations such as this presents no problem in team theory, as information structures are perfectly capable of modeling uncertain information sources. The hard questions that arise are how to structure the pooling of information between sensors with dependent information, how to take action in the face of uncertainty, and what control methods are most appropriate for directing the gathering of information. We are currently exploring these issues.

## 4.2 Loss Considerations and Control

The loss function associated with an agent or a team determines the essential nature of the decision maker. In the standard optimal control formulation, specification of information structures and loss provide the criteria for selection of the *optimal control law or decision rule*. However, optimal rules are often difficult to derive, and have a computationally complex nature. General results are known only for a restricted class of information structure/loss function formulations. Another method for selecting controls is to postulate as class of *admissible controls*, and choose the member of this class which minimizes the loss. Lastly, we can consider constructing decision rules *ad hoc* and evaluating their performance relative to an objective based on simulation studies. In any case, the character of the loss function is crucial in determining the resultant decision rule or control law.

One area which needs more exploration is a methodology for the specification of loss functions. Ideally, the loss function should be justifiable in terms of objective criteria related to the problem. Pragmatically, it is often dictated by mathematical convenience. From the team perspective, more work needs to be done on the interaction of team and local loss characterizations. Section 2 presented some results in this direction, but more work is surely needed, particularly in the case where the team objectives are not expressible as some combination of member's objectives.

To illustrate what we have in mind, consider formulating loss functions for controlling a system based on a desired state of information. That is, if the team has its goal some state of information (for example, to move an arm this information is needed), what action is most appropriate for progressing from the current information state toward the desired information state. Should it select an action which will change the uncertainty associated with current information, or go ahead with an action that adds uncorrelated evidence? How should it decide that it has enough information? More concretely, should the executive take another picture with the camera, or perhaps take a different view, or maybe use another sensor altogether. Maybe the sensors themselves should decide individually what to do. These are all issues dealing with the interaction of information and action. By using team theory, we can easily formulate the problem, specify loss functions or decision methods based on, for example, the parameters of a probability distribution associated with some information source, and examine the results via simulation or by analytic methods.

### 4.3 Decision Theory and AI

As we stated at the outset, we consider our work relevant to AI in that we that we may want to consider information as interpreted, and would like to consider parts of a system as intelligent reasoning agents. In related work dealing with the interaction of intelligent agents, Rosenschein and Genesereth in [22] and Ginsburg in [9] have investigated variations on game theoretical definitions of rationality for coordinating intelligent agents. However these results are an attempt to analyze the interaction of intelligent agents with *no a priori structure* and investigate the consequence of various rationality assumptions. We, on the other hand, postulate a given team structure and are interested in discovering its properties. This is an important fundamental distinction to keep in mind.

It is our view that knowledge-based reasoning agents can be used effectively in the team theoretic framework; but, we must be able to describe them in terms of the other systems elements — that is, as decision makers with information structures and preferences about action. In order to achieve this objective, we must develop information structures to be compatible with AI conceptions of information as discrete (usually logical) tokens, and somehow connect control structures and loss formulations. At his point, we can sketch at least possibility. First, view such reasoning agents as consisting of two phases: computing the information structure, and selecting an optimal action in the face of available information. This is similar to the classic separation of estimation in control in control theory literature[3]. Computation of the information structure amounts to using furnished information and making implicit information explicit relative to a given model[23]. That is, some part of the information in the knowledge base is used to infer new facts from given information. The complete set of such facts form (in a limiting sense) the information structure. Some of the theoretical analyses of (logical) knowledge have detailed methods for describing this process of inference using variants of modal logic[23].

Loss formulations for the preference of actions can be specified using a conception of action similar to the situation calculus[17,19]. In this system, action is a mapping between world states, where each state represents a configuration of the world. Moore [19] has shown how both information and action can be represented and related within the conceptual framework of world states, making loss formulations based on information possible. The actual details of this procedure are beyond the scope of this paper, but we can show that several problems in the planning domain can, in fact, be reduced to decision problems posed in this manner. As a further example, consider building a decision-maker who attempts to fill in gaps in an incomplete, discrete knowledge base. The specification of information and loss functions can be done in terms of world states as presented above, and the actual implementation of the system done as a rule-based system.

Finally, we may attempt to combine this agent with agents which attempt to reduce uncertainty in the probabilistic sense outlined in the previous subsection. For instance, a camera and a tactile sensor which have local probabilistic uncertainty reduction methods, and a global executive which is building models of the environment. Using team theory, we can analyze possible methods for control and cooperation of these disparate agents and offer a coherent explanation of the full system's behavior.

## 5 Conclusions and Future Research

Analysis of the general team organization with respect to team members information structures provides a systematic framework for addressing a number of important questions concerning the effect of sensor agent capabilities on overall system performance. We summarize some of the more important issues:

1. Could sensor benefit by guidance from another team member. Should communication between members be increased.

2. Should the sensors ability to make observations be enhanced in anyway, by changing hardware or finding algorithmic bottlenecks.

3. When would an exchange of opinions and dynamic consensus be attempted.

4. What overall system structure (as described by the information structures of the team members) is best (or better) for different tasks.

Similarly, there are a number of important questions that can be addressed by analyzing the effect of individual team members utility and decision functions, including:

1. Communication and time costs in the decision process to provide for real time action.

2. Inclusion of decisions to take new observations of the environment if previous opinions are rejected by other team members, or if insufficient information was obtained on a first pass.

3. Effects of new decision heuristics on overall system performance.

Of course all these ideas may well be difficult to consider analytically, though this formalism does reduce the search space of alternatives and provides a framework within which these issues may be evaluated. The team theoretic organization is a powerful method for analyzing multi-agent systems, but it is certainly not the complete answer.

**Acknowledgment:** The Authors would like to thank Dr. Max Mintz for many valuable discussions about this subject.